\documentclass[11pt]{article}

\input{setup.sty}

\title{Constrained HRT Surfaces and their Entropic Interpretation}

\author[1]{Xi Dong,} \emailAdd{xidong@ucsb.edu}

\author[1]{Donald Marolf,} \emailAdd{marolf@ucsb.edu}

\author[1,2]{and Pratik Rath} \emailAdd{pratik\_rath@berkeley.edu}

\affiliation[1]{Department of Physics, University of California, Santa Barbara, CA 93106, USA}
\affiliation[2]{Center for Theoretical Physics and Department of Physics,
University of California, Berkeley, CA 94720, USA}

\abstract{Consider two boundary subregions $A$ and $B$ that lie in a common boundary Cauchy surface, and consider also the associated HRT surface $\gamma_B$ for $B$.  In that context, the constrained HRT surface $\gamma_{A:B}$ can be defined as the codimension-2 bulk surface anchored to $A$ that is obtained by a maximin construction restricted to Cauchy slices containing  $\gamma_B$. As a result, $\gamma_{A:B}$ is the union of two pieces, $\gamma^B_{A:B}$ and $\gamma^{\bar B}_{A:B}$ lying respectively in the entanglement wedges of $B$ and its complement $\bar B$.
Unlike the area $\cA\(\gamma_A\)$ of the HRT surface $\gamma_A$, at least in the semiclassical limit, the area  $\cA\(\gamma_{A:B}\)$ of $\gamma_{A:B}$ commutes with the area $\cA\(\gamma_B\)$ of $\gamma_B$.
To study the entropic interpretation of $\cA\(\gamma_{A:B}\)$, we analyze the R\'enyi entropies of subregion $A$ in a fixed-area state of subregion $B$.  We use the gravitational path integral to show that the $n\approx1$ R\'enyi entropies are then computed by minimizing $\cA\(\gamma_A\)$ over spacetimes defined by a boost angle conjugate to $\cA\(\gamma_B\)$.
In the case where the pieces $\gamma^B_{A:B}$ and $\gamma^{\bar B}_{A:B}$ intersect at a constant boost angle, a geometric argument shows that the $n\approx1$ R\'enyi entropy is then given by $\frac{\mathcal{A}(\gamma_{A:B})}{4G}$.
We discuss how the $n\approx1$ R\'enyi entropy differs from the von Neumann entropy due to a lack of commutativity of the $n\to1$ and $G\to0$ limits.
We also discuss how the behaviour changes as a function of the width of the fixed-area state.
Our results are relevant to some of the issues associated with attempts to use standard random tensor networks to describe time dependent geometries.}

\begin{document}

\maketitle

\section{Introduction}

\label{sec:introduction}

In the context of the AdS/CFT correspondence \cite{Maldacena_1999}, the connection between entanglement measures in the boundary theory and areas of surfaces in the bulk theory has been an active area of research.
The Ryu-Takayanagi formula \cite{Ryu:2006bv}, and its generalization to time dependent spacetimes by Hubeny-Rangamani-Takayanagi (HRT) \cite{Hubeny:2007xt}, led the way in establishing a duality between the von Neumann entropy of boundary subregion $R$ and the area of an extremal surface $\gamma_R$ anchored to $\partial R$.
Here and in the main text below we restrict discussion to the case where the bulk gravity theory is described by the Einstein-Hilbert action with minimally-coupled matter.  The AdS/CFT dictionary has also been extended to many more entanglement measures including the R\'enyi entropy \cite{Headrick:2010zt,Hung:2011nu,Hartman:2013mia,Faulkner:2013yia,Lewkowycz:2013nqa,Dong:2016fnf}, entanglement negativity \cite{Kudler-Flam:2018qjo, Kusuki:2019zsp, Dong:2021clv} and reflected entropy \cite{Dutta:2019gen,Akers:2021pvd,Akers:2022max,Akers:2022zxr} among others.

\begin{figure}
    \centering
    \includegraphics[scale=0.3]{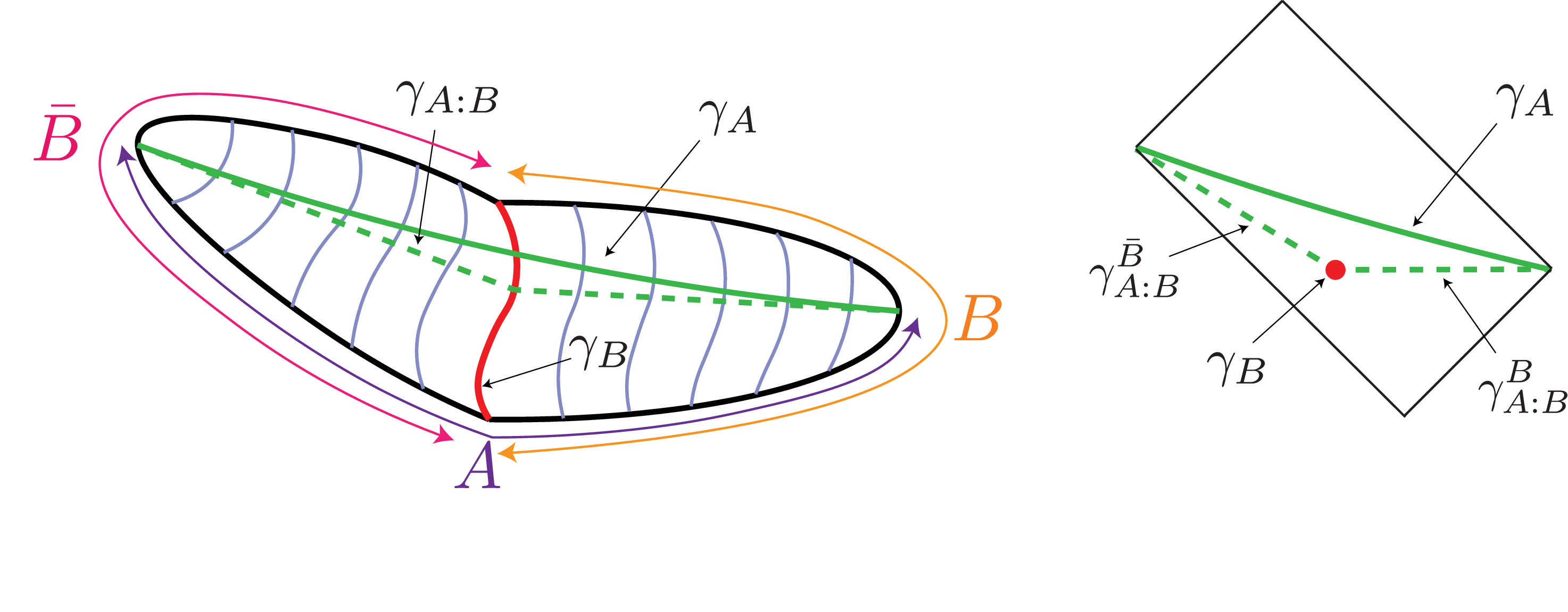}
    \caption{Left: The HRT surface $\gamma_B$ (red) for subregion $B$ (marked with a long orange double arrow), along with a family of spacelike separated HRT surfaces (light blue) corresponding to nested subregions can be placed on the same Cauchy slice. The HRT surface $\gamma_A$ (solid, green) for subregion $A$ (marked with a long purple double arrow) does not generally lie on the same Cauchy slice. The constrained HRT surface $\gamma_{A:B} = \gamma^B_{A:B} \cup \gamma^{\bar B}_{A:B} $ (dashed, green) is extremal everywhere except at the location where it meets $\gamma_B$. Right: A side view of the same configuration.}
    \label{fig:cHRT}
\end{figure}

An interesting such proposal was made in Ref.~\cite{Chen:2018rgz}, which considered a surface that we call the constrained HRT surface $\gamma_{A:B}$.  This
$\gamma_{A:B}$ is defined when $A$, $B$ are boundary subregions that lie on a common boundary Cauchy surface.  In that case, the codimension-2 surface $\gamma_{A:B}$
is defined by minimizing the area of spacelike\footnote{While the spacelike qualifier was not explicitly stated in Ref.~\cite{Chen:2018rgz}, it is implied by the idea that the area of the relevant surfaces defines a partial order with respect to which one can minimize.} surfaces anchored to $\partial A$ (and appropriately homologous to $A$) that are extremal everywhere except at possible intersections with the HRT surface $\gamma_B$ for subregion $B$;  see \figref{fig:cHRT}.  It was also suggested in Ref.~\cite{Chen:2018rgz} that, at least under appropriate conditions,  no point on $\gamma_{A:B}$ can be timelike separated from any point on $\gamma_B$ (and we will show this below under the assumptions of Ref.~\cite{Wall:2012uf}).  As a result, $\gamma_B$ defines two pieces of $\gamma_{A:B}$ which we will call $\gamma^B_{A:B}$ and $\gamma^{\bar B}_{A:B}$, that respectively lie in the entanglement wedges\footnote{We define the entanglement wedges to be closed, and in particular include $\gamma_B$.} of $B$ and its complement $\bar B$.
Ref.~\cite{Chen:2018rgz} proposed that when the intersection between $\gamma^B_{A:B}$ and $\gamma^{\bar B}_{A:B}$ defines a constant boost angle, the quantity $\frac{\cA\(\gamma_{A:B}\)}{4G}$ computes the von Neumann entropy of $A$ minimized over modular flow applied to subregion $B$. 
In this paper, we revisit constrained HRT surfaces finding a different (but closely related) entropic interpretation for their area.

In \secref{sec:commute}, we provide a maximin construction of the constrained HRT surface.
We also note that $\{\mathcal{A}\(\gamma_{A:B}\),\mathcal{A}\(\gamma_{B}\)\}=0$, where $\mathcal{A}$ denotes the area of the specified codimension-2 surface and $\{f,g\}$ is the Poisson/Peierls bracket of observables $f$ and $g$.
This demonstrates that the area operators of $\gamma_{A:B}$ and $\gamma_B$ commute in the classical limit.\footnote{To define an area operator carefully in the quantum theory, we need to impose a bulk ultraviolet cutoff to suppress divergences coming from quantum fluctuations.}

We then move on to discuss the entropic interpretation of $\cA\(\gamma_{A:B}\)$.  As shown in \figref{fig:cHRT}, the tangent spaces of $\gamma^B_{A:B}$ and $\gamma^{\bar B}_{A:B}$ may not be continuous across $\gamma_B$. We now characterize this potential discontinuity by defining a boost angle. The minimization over possible intersections with $\gamma_B$ in the definition of $\cA\(\gamma_{A:B}\)$ implies that the projections of those tangent spaces onto directions parallel to $\gamma_B$ are actually continuous across $\gamma_B$. So at any given intersection point where $\gamma^B_{A:B}$ meets $\gamma^{\bar B}_{A:B}$, we only need to consider the projections of their tangent spaces onto the timelike plane orthogonal to $\gamma_B$. As long as these two projections are nonzero\footnote{There are special situations where these projections vanish, such as when a connected component of $\gamma_{A:B}$ coincides with a connected component of $\gamma_B$. At these special intersection points, the boost angle is not defined.\label{coincide}} and thus are one-dimensional vector spaces, they define a boost angle that relates one to the other.
Like the conjecture of Ref.~\cite{Chen:2018rgz}, our proposal considers the case where the boost angle is the same at all intersection points where the boost angle is defined. It then states that, for $\epsilon=O(G^0)$ but small and positive, 
 the $n=1+\epsilon$ R\'enyi entropy of subregion $A$ in a fixed-area state of subregion $B$ is given by
\begin{equation}\label{eq:Renyi}
    S_n(A) = \frac{\mathcal{A}(\check \gamma_{A:B})}{4G} +O\(\frac{\epsilon}{G}\).
\end{equation}
Here and below the use of a check in denoting a constrained HRT surface (as in $\check \gamma_{A:B}$) indicates that the boost angle between the pieces $\gamma^B_{A:B}$ and $\gamma^{\bar B}_{A:B}$ is in fact independent of the intersection point.\footnote{Ref.~\cite{Chen:2018rgz} instead used a superscript $^>$ for the constant boost-angle case.}

In order to obtain the result \Eqref{eq:Renyi}, we will first discuss the calculation of entropies in fixed-area states in \secref{sec:entropic}.
In particular, it was argued in Ref.~\cite{Kaplan:2022orm} that the area ${\cal A}(\gamma_B)$ of an 
HRT-surface $\gamma_B$ generates a flow on the covariant phase space that acts as 
boundary-condition-preserving (BCP) kink transformation which, as described in Ref.~\cite{Bousso:2020yxi}, inserts a relative boost between the entanglement wedges of $B$ and $\bar B$.   As will be discussed in detail below, this means that a fixed-area state for $\gamma_B$ must decompose into a superposition of states with different relative boosts across $\gamma_B$.
Furthermore, when the area is sharply-fixed (so as to define a pseudo-eigenstate in the language of Ref.~\cite{Dong:2022ilf}),  the superposition will explore a large range of such relative boosts.

For such a pseudo-eigenstate, and assuming replica symmetry a la Lewkowycz-Maldacena \cite{Lewkowycz:2013nqa}, we consider the limit of small $\epsilon$ in which backreaction from the cosmic brane \cite{Dong:2016fnf} at the HRT surface can be neglected.
The associated path integral is then dominated by the choice of boost angle $s$ that minimizes the area of the associated HRT surface $\gamma_A(s)$.
In the case where the pieces $\gamma^B_{A:B}$, $\gamma^{\bar B}_{A:B}$ of $\gamma_{A:B}$ intersect at constant boost angle,  a maximin argument along the lines of Ref.~\cite{Wall:2012uf} shows that the dominant saddle is the one where the HRT surface $\gamma_A(s)$ agrees with the constrained HRT surface $\gamma_{A:B}$, thus establishing \Eqref{eq:Renyi}.

In the above context, it is important to distinguish the $n\approx1$ R\'enyi entropy from  the von Neumann entropy.  The two turn out to behave differently, thus showing that the $n\to 1$ and $G\to 0$ limits do not commute.
While the $n\approx1$ R\'enyi entropy is computed by $\min_s \frac{\cA\(\gamma_A(s)\)}{4G}$, the von Neumann entropy is given by the expectation value $\langle \frac{\cA\(\gamma_A(s)\)}{4G}\rangle$, where the average is performed over $s$ and is weighted by the probability distribution defined by the gravitational path integral.

In \secref{sec:disc}, we discuss the implications of our work for tensor networks.
Moreover, we discuss the potential addition of higher-derivative and quantum corrections to our analysis.


\section{Constrained HRT surfaces}
\label{sec:commute}

Consider two boundary subregions $A$ and $B$ that can be placed on the same boundary Cauchy slice.
If $A\cap B\neq \emptyset$, 
the corresponding HRT surfaces $\gamma_{A}$ and $\gamma_B$ generally contain portions that are timelike separated from each other; see again \figref{fig:cHRT}. 
As discussed in e.g. Ref.~\cite{Bao:2018pvs}, this suggests that the areas of these surfaces will fail to commute with each other in a quantum theory of gravity, i.e., $\[\mathcal{A}\(\gamma_A\),\mathcal{A}\(\gamma_B\)\]\neq 0$; see e.g. Ref.~\cite{Kaplan:2022orm} for explicit semiclassical examples of such non-vanishing commutators.
As a result, we cannot simultaneously project onto fixed eigenvalues of the two operators.
The classical limit of the above statement is that the Poisson/Peierls bracket between the two observables is non-vanishing, i.e., $\{\mathcal{A}\(\gamma_A\),\mathcal{A}\(\gamma_B\)\}\neq 0$.

In contrast, Ref.~\cite{Chen:2018rgz} defined a constrained analogue of an HRT surface that we will call $\gamma_{A:B}$.
 This $\gamma_{A:B}$ was defined as the codimension-2 surface anchored to $\partial A$ (and appropriately homologous to $A$) 
 that has minimal area among all such surfaces that are extremal everywhere except at potential intersections with $\gamma_B$.  
 It was also suggested in Ref.~\cite{Chen:2018rgz} that, at least under appropriate conditions,  no point on $\gamma_{A:B}$ can be timelike separated from any point on $\gamma_B$.

This motivates us to provide a restricted maximin definition (inspired by the maximin construction of Ref.~\cite{Wall:2012uf}) for the area ${\cal A}(\gamma_{A:B})$ wherein we restrict to Cauchy slices that include all of $\gamma_B$ as well as the boundary anchors defined by $A$.  In doing so, we assume as in Ref.~\cite{Wall:2012uf} that our spacetime satisfies the null energy condition, and that bulk Cauchy surfaces $\Sigma$ for which $\partial\Sigma$ is a given Cauchy surface of the asymptotically locally AdS boundary can touch future and past boundaries only at Kasner-like singularities.\footnote{If this last assumption is dropped, then HRT surfaces may fail to exist, or they may fail to correctly compute the desired entropies. See Ref.~\cite{Fischetti:2014uxa} for discussion and proposals for dealing with such situations.}

That such Cauchy slices exist is clear from e.g. the maximin construction of $\gamma_B$ given in Ref.~\cite{Marolf:2019bgj}.
On any Cauchy slice $\Sigma_B \supset \gamma_B$, we first find the minimal area surface anchored to $\partial A$.
We then maximize over the choice of $\Sigma_B$  to obtain a maximin surface $\gamma^{mm}_{A:B}$.  We assume as in Ref.~\cite{Wall:2012uf} that the associated stability criterion is satisfied. When the relevant Cauchy surface is both spacelike and smooth, this can be shown using the technical argument in Sec.~3.5 of Ref.~\cite{Wall:2012uf}, but it remains an assumption more generally.  With this assumption, the surface $\gamma^{mm}_{A:B}$ exists and can be studied as in Ref.~\cite{Wall:2012uf}.

The usual arguments are then readily applied near points of $\gamma^{mm}_{A:B}$ located away from $\gamma_B$, showing that $\gamma^{mm}_{A:B}$ is extremal at such points. Let us refer to the class of all such extremal-away-from-$\gamma_B$ surfaces (including those that do not intersect $\gamma_B$ at all) as ${\cal E}_{\gamma_B}$, and let us also recall that $\gamma_{A:B}$ as defined in Ref.~\cite{Chen:2018rgz} is the minimal-area surface\footnote{In \cite{Chen:2018rgz} it was simply assumed that such a minimum exists.  But we will soon see that  $\gamma^{mm}_{A:B}$ must be the minimum, so a minimum does in fact exist.} in ${\cal E}_{\gamma_B}$.     Thus we obtain the bound
\begin{equation}
\label{eq:mmG}
{\cal A}(\gamma_{A:B}) \le {\cal A}(\gamma^{mm}_{A:B}).
\end{equation}

We will now also show that $\gamma^{mm}_{A:B}$  must minimize the area within ${\cal E}_{\gamma_B}$ by using the concept of the representative of $\gamma_{A:B}$ on an arbitrary Cauchy surface containing $\gamma_B$, and in particular using the Cauchy surface $\Sigma$ on which $\gamma^{mm}_{A:B}$ is minimal.  This representative was defined in Sec.~3.1 of Ref.~\cite{Wall:2012uf} as the intersection of $\Sigma$ with a null congruence orthogonal to $\gamma_{A:B}$.  The focussing argument of Ref.~\cite{Wall:2012uf} again shows that the area of the representative $r(\gamma_{A:B})$ on $\Sigma$ can be no larger than the area of $\gamma_{A:B}$.  And since $\gamma^{mm}_{A:B}$ is minimal on $\Sigma$, we must also have
\begin{equation}
\label{eq:mmL}
\cA (\gamma^{mm}_{A:B}) \le \cA (r(\gamma_{A:B})) \le \cA (\gamma_{A:B}).
\end{equation}
Comparing \Eqref{eq:mmG} and \Eqref{eq:mmL} shows that the areas of $\gamma^{mm}_{A:B}$ and $\gamma_{A:B}$ must coincide.   This will generically also imply that $\gamma^{mm}_{A:B}$ and $\gamma_{A:B}$ define the same surface, though equality of their areas will be enough for our purposes.  In particular, since this equality holds in all spacetimes that satisfy the equations of motion, the areas $\cA(\gamma_{A:B})$ and $\cA(\gamma^{mm}_{A:B})$ define identical functions on the phase space of classical solutions.  Focussing on the areas in this way allows a clean way to deal with cases where the minimal surface is not unique.  Indeed, in such non-unique cases we are always free to choose $\gamma_{A:B} = \gamma^{mm}_{A:B}$, which we will do below.  

For later purposes, we also note that our maximin construction yields the bound
\begin{equation}
\label{eq:cHRTlb}
\cA (\gamma_{A:B}) \le \cA(\gamma_A).
\end{equation}
This follows directly from the fact that $\gamma_A$ can be obtained from the similar-but-less-restrictive maximin principle stated in Ref.~\cite{Wall:2012uf}.  In particular, since $\gamma_A$ is obtained by maximizing the area over a strictly larger set of Cauchy surfaces, it must be at least as large as our restricted maximum $\cA (\gamma_{A:B})$.

The restricted maximin construction above establishes that the surfaces $\gamma_B$ and $\gamma_{A:B}$  lie in a common Cauchy slice. This opens up the possibility that $\{\mathcal{A}(\gamma_{A:B}),\mathcal{A}(\gamma_{B})\}=0$. 
To see that this is indeed the case, we recall from Ref.~\cite{Kaplan:2022orm} that, in Einstein-Hilbert gravity,  the flow generated by $\mathcal{A}(\gamma_{B})$ is given by a BCP kink-transformation.
This transformation has a simple description on any Cauchy slice $\Sigma_B$ that includes $\gamma_B$. While the induced metric is invariant under the transformation, the extrinsic curvature shifts according to\footnote{Our flow parameter $s$ corresponds to $-s$ in the convention of Ref.~\cite{Kaplan:2022orm}.}
\begin{align}\label{eq:kink}
        K_{\perp \perp}&\rightarrow K_{\perp \perp} + 2\pi s \,\delta_{\gamma_B}.
\end{align}
Here $\delta_{\gamma_B}$ is a Dirac delta function localized on $\gamma_B$ and  $\perp$ represents the normalized vector tangent to $\Sigma_B$ but orthogonal to $\gamma_B$; see Ref.~\cite{Kaplan:2022orm} for details.
The initial data on $\Sigma_B$ can then be used to obtain the corresponding spacetime by solving the equations of motion.

In particular, the causal development of the data on $\Sigma_B$ on either side of $\gamma_B$ is unaffected by the transformation (since the change in the initial data on $\Sigma_B$ is localized to $\gamma_B$).   Furthermore, since $\gamma_{A:B}$ is nowhere timelike-separated from $\gamma_B$, away from $\gamma_B$ it must be entirely contained in that causal development.  
In certain cases, there can also be a finite contribution to the area of $\gamma_{A:B}$ from its intersection with $\gamma_B$ (see footnote~\ref{coincide}), but this area will be determined by the induced metric on any Cauchy slice that contains $\gamma_B$, and this induced metric is also invariant under the transformation.  It thus follows that our flow cannot change ${\cal A} (\gamma_{A:B})$, and thus that the Poisson/Peierls bracket  $\{\mathcal{A}(\gamma_{A:B}),\mathcal{A}(\gamma_{B})\}$ must vanish.  At the quantum level, the two areas will then form a commuting set of operators at least at leading order in $\hbar$.

\section{Entropic Interpretation}
\label{sec:entropic}

We now analyze the entropic interpretation of the constrained HRT surface $\gamma_{A:B}$.
In order to do so, we will consider the calculation of the R\'enyi entropy of subregion $A$ in a state where the area of the HRT surface $\gamma_B$ is fixed polynomially sharply (i.e., the width is polynomially small in $G$).
Using the gravitational path integral, we will explain how the fixed-area state decomposes into a superposition over BCP kink-transformed states.\footnote{At the quantum level we simply define the BCP kink transform to be the transformation generated by exponentiating the HRT area $\cA(\gamma_B)$.}
In the case when the intersection between $\gamma_{A:B}$ and $\gamma_B$ happens at a constant boost angle, we will show that $\frac{\cA(\gamma_{A:B})}{4G}$ computes the $n\approx 1$ R\'enyi entropy of subregion $A$ in the fixed-area state of subregion $B$.
We will also clarify the difference between the $n\approx1$ R\'enyi entropy and the entanglement entropy, highlighting an order-of-limits issue which obstructs us from interpreting the area of the constrained HRT surface as the von Neumann entropy of a subregion in our fixed-area state.

Consider a CFT state $\ket{\psi}$ whose norm is computed by a Euclidean path integral with smooth sources on a manifold $\mathcal{M}$.
The corresponding bulk state is prepared using a gravitational path integral with the appropriate smooth boundary conditions, i.e.,
\begin{equation}
\label{eq:GPINpsi}
    \bra{\psi}\ket{\psi} = \int_{\partial \mathcal{B}=\mathcal{M}} Dg\, e^{-I[g]}.
\end{equation}
In the saddle point approximation, \Eqref{eq:GPINpsi} will be dominated by a spacetime  which satisfies the Einstein equations everywhere and which also satisfies the boundary condition $\partial \mathcal{B}=\mathcal{M}$.
To obtain a fixed-area state for subregion $B$, we simply restrict this state to a small window of areas; i.e.,
 we construct
 \begin{equation}
 \label{eq:projstate}
    \ket{\psi_{\cA_0}} = \Pi_{\cA_0} \ket{\psi},
\end{equation}
where $\Pi_{\cA_0}$ is an operator that sharply restricts the wavefunction to eigenvalues of $\cA(\gamma_B)\equiv\cA_B$ around $\cA_0$. 

So long as the above window contains a large number of area-eigenvalues, there will be an approximate notion of a `conjugate' variable to $\cA_B$.  Semiclassically, this conjugate is the relative boost angle $s_B$ across $\gamma_B$ \cite{Kaplan:2022orm,Dong:2022ilf} (see also Refs.~\cite{Jafferis:2015del,Bousso:2020yxi,Witten:2021unn,Chandrasekaran:2022eqq}), i.e., we have the uncertainty relation
\begin{equation}
    \Delta \(\frac{\cA_B}{4G}\) \Delta s_B \gtrsim O(1).
\end{equation}
For semiclassical states prepared using a (say, Schwinger-Keldysh) gravitational path integral, we typically have $\Delta \cA_B,\Delta s_B =O(\sqrt{G})$.
Reducing the variance in $\cA_B$ increases the variance in $s_B$.  However, as long as both $\Delta \cA_B$ and $\Delta s_B$ vanish in the $G\to 0$ limit, the limit is still associated with a well defined classical geometry.  In particular, 
the Lorentzian spacetimes associated to such squeezed states were discussed in Ref.~\cite{Dong:2022ilf}.

On the other hand, for $\Delta \cA_B$ of order $G$ or smaller, the corresponding $\Delta s_B$ fail to vanish  as $G\to0$ and the state is not semiclassical.
We will focus on states of this sort, with the restriction that $\Delta \cA_B$ vanish no faster than polynomially as $G\to 0$.  Thus we still expect the associated density matrix for each region $B, \bar B$  to have an exponentially large number of eigenvalues.
Such states were called pseudo-eigenstates in Ref.~\cite{Dong:2022ilf}.
These are precisely the kind of states where we will be able to provide an entropic interpretation for $\gamma_{A:B}$.

To be specific, we can choose to define the operator $\Pi_{\cA_0}$ by using a Gaussian window function
\begin{equation}
\label{eq:pseudoproj}
\Pi_{\cA_0}=f(\hat{\cA}_B) = N \exp\[-\frac{(\hat{x}_B-x_0)^2}{2\sigma^2} - i \hat{x}_B s_0\],
\end{equation}
where $N$ is a normalization factor, $x=\frac{\cA}{4G}$, and we take $s_0$ independent of $G$.  We will correspondingly refer to $\Pi_{\cA_0}$ as a {\it window operator}.  
Note that we can also write \Eqref{eq:pseudoproj} in the form
\begin{equation}\label{eq:op}
    f(\hat{\cA}_B) = N' \int ds \,\tilde{f}(s) \exp\[-i \hat{x}_B s\],
\end{equation}
where $\tilde{f}(s)$ is the Fourier transform of $f$ (up to normalization), i.e.,
\begin{equation}
\label{eq:fsGauss}   
    \tilde{f}(s) = \exp\[-\frac{\sigma^2}{2}(s-s_0)^2+i x_0 (s-s_0)\].
\end{equation}
As described above, the desired pseudo-eigenstates \eqref{eq:pseudoproj} have $\sigma=O(1)$ or parametrically smaller in $G$.  We will choose to focus on $\sigma=O(G^\alpha)$ with some $\alpha>0$. As a result, the width of \Eqref{eq:fsGauss} is polynomially large in $1/G$.
We also see that the parameter $s_0$ determines the peak of the probability distribution for the boost angle.

\begin{figure}
    \centering
    \includegraphics[scale=0.2]{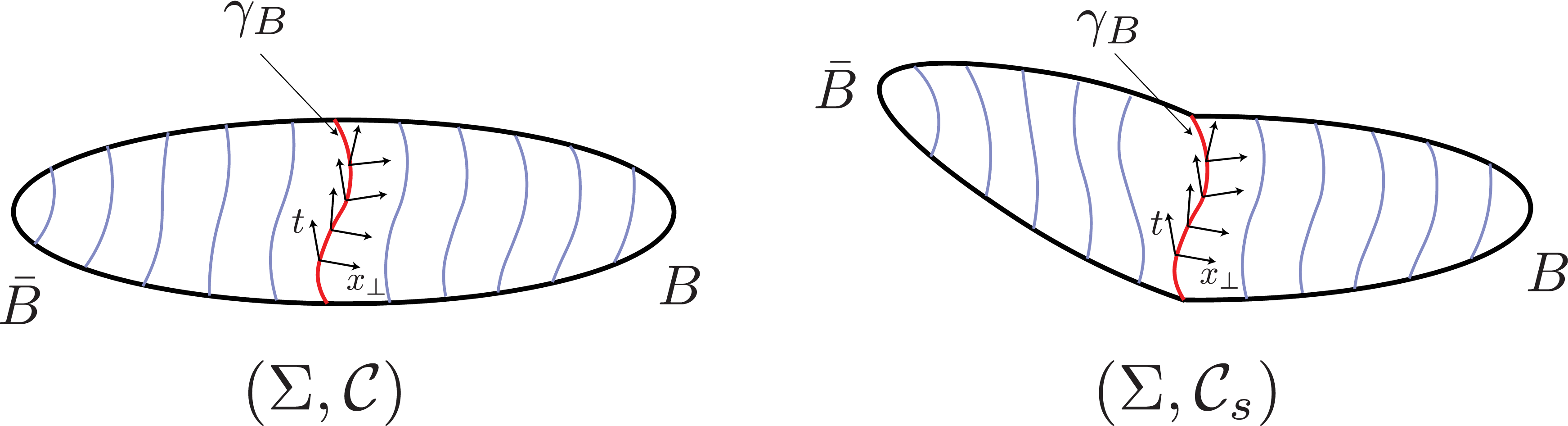}
    \caption{The boundary-condition-preserving (BCP) kink-transformation modifies the Cauchy data ${\cal C}$ on a bulk Cauchy slice $\Sigma$ to new data ${\cal C}_s$ by adding a delta function in the codimension-1 extrinsic curvature component $K_{\perp \perp}$, where as shown $x_{\perp}$ represents the intrinsic coordinate that measures proper distance on $\Sigma$ in the direction orthogonal to the HRT surface $\gamma_B$.
    This essentially results in a shift in the relative boost $s$ across $\gamma_R$. The qualifier BCP emphasizes that the slice $\Sigma$ with the new data ${\cal C}_s$ remains glued to the asymptotic boundary regions $B$ and $\bar{B}$ in precisely the same way as for $\Sigma$ with the original data ${\cal C}$.}
    \label{fig:BCP}
\end{figure}

\begin{figure}
    \centering
    \includegraphics[scale=0.5]{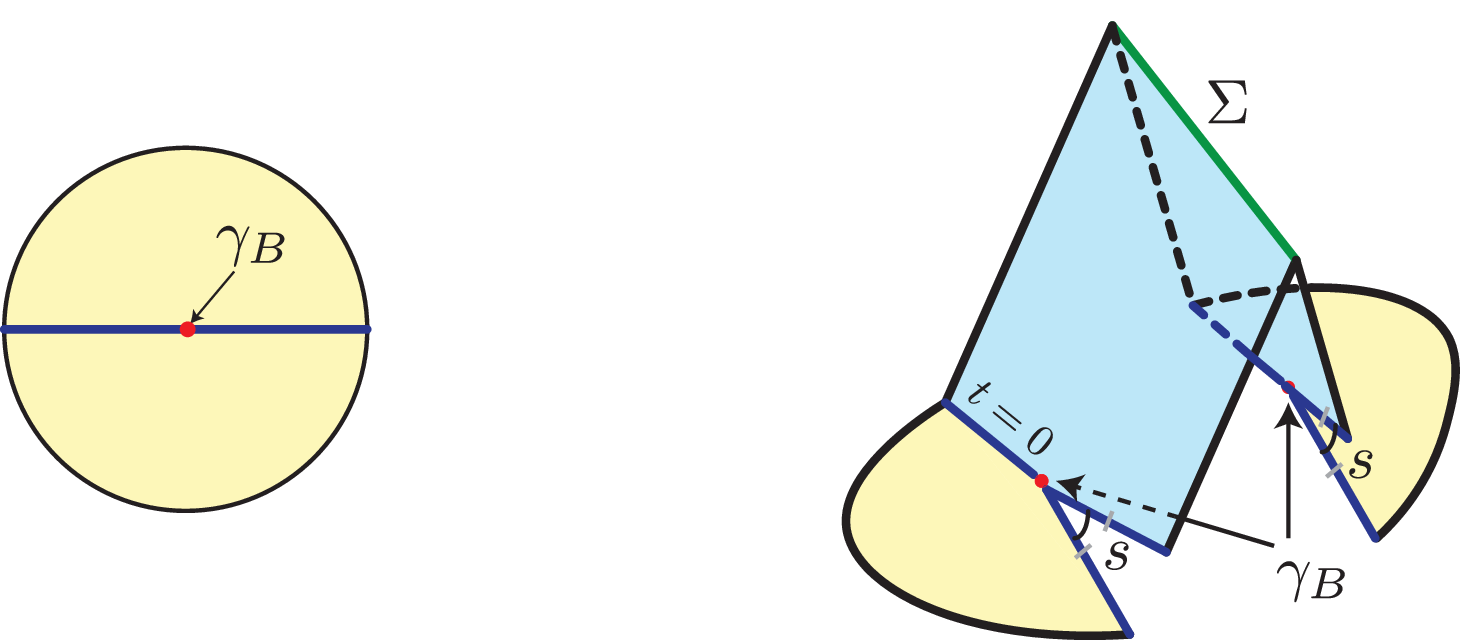}
    \caption{{\bf Left:} A saddle for the (Euclidean or complex) path integral for correlators in $\ket{\psi}$.  The blue line indicates a surface with initial data that defines the associated Lorentz-signature solution and which contains $\gamma_B$. The same Euclidean saddle computes the norm of any $|\psi(s)\rangle$, since they are related by unitary operators. {\bf Right:}  A Schwinger-Keldysh contour that attaches a piece of Lorentz-signature spacetime at the blue surface in fact yields two copies of $\gamma_B$.  For any fixed value of $s$, in the semiclassical approximation, the path integral for correlators in $\ket{\psi(s)}$ inserts the operator $e^{-i\hat x_B s}$ at both copies, enacting a relative boost of rapidity $s$ between the initial data on the half of the blue slice on the closer side of $\gamma_B$ and the data on the farther side.  The Lorentzian region (light blue) is then a BCP kink-transform of the corresponding region for the original $\ket{\psi}$.}
    \label{fig:lor}
\end{figure}

Since the state $\ket{\psi}$ is defined by a gravitational path integral with smooth boundary-sources, in the semiclassical limit it will be associated with some spacetime geometry $g_\psi$.  When we apply \Eqref{eq:op} to $\ket{\psi}$, we will obtain an integral over states $\ket{\psi(s)} := e^{-i\hat x_B s}\ket{\psi}$.  These states will also be semiclassical and will be associated with spacetime geometries $g_{\psi(s)}$ obtained by applying the boundary-condition-preserving (BCP) kink transformation to $g_\psi$ as summarized in \figref{fig:BCP}.  This can be seen, for example, from the fact that the insertion of $e^{-i\hat x_B s}$ into the gravitational path integral is equivalent to including a cosmic brane that classically sources a Lorentz-signature conical defect.  As a result, Schwinger-Keldysh path integrals that compute correlators in the state $\ket{\psi(s)}$ can be approximated by saddles that agree with the saddle for $\ket{\psi}$ near the Euclidean boundaries, but which contain new Lorentz-signature conical defects at $\gamma_B$.  A Schwinger-Keldysh path integral will in fact contain two copies of $\gamma_B$, and the desired correlators are computed by inserting operators in the region between the resulting conical singularities; see \figref{fig:lor}.  As a result, at the semiclassical level, those correlators probe the spacetime $g_{\psi(s)}$ constructed by applying the BCP kink transformation to $g_\psi$ as claimed above.

The Lorentzian spacetimes $g_{\psi(s)}$ are well understood \cite{Bousso:2020yxi}.
In Einstein gravity, the BCP kink transformed initial data results in Weyl curvature delta function shocks along the lightcones of the HRT surface but the rest of the spacetime is expected to be smooth.

Using \Eqref{eq:op}, the fixed-area state can be written as
\begin{align}\label{eq:dis}
    \ket{\psi_{\cA_0}}&= N' \int ds \tilde{f}(s) \ket{\psi(s)}.
\end{align}
Correlation functions in the future/past of $\gamma_B$ can be computed by the expression
\begin{equation}\label{eq:cor}
\bra{\psi_{\cA_0}}O\ket{\psi_{\cA_0}}= |N'|^2 \int ds\,ds'\, \tilde{f}(s)\,\tilde{f}^*(s')\, \langle \psi(s')|O|\psi(s)\rangle.
\end{equation}
It is convenient to rewrite the flow parameters as $\bar{s}=\frac{s+s'}{2}$ and $\Delta=\frac{s-s'}{2}$. The overlap $\bra{\psi(\bar{s}-\Delta)}\ket{\psi(\bar{s}+\Delta)}$ may be evaluated using a Euclidean gravitational path integral, and for a given $\bar s$ the Cauchy-Schwarz inequality makes clear that the overlap is maximized at $\Delta=0$. And since the overlap is computed by a semiclassical gravitational path integral, away from $\Delta=0$ it will decay as $\exp\[-\frac{\#\Delta^2}{G}\]$.\footnote{To the extent that we may replace area-flow by modular flow we may write $|\bra{\psi(\bar{s}+\Delta)}\ket{\psi(\bar{s}-\Delta)}|^2=\left|\tr\big(e^{-(1+2i\Delta) K_B^\psi}\big)\right|^2$ and note that this quantity can be thought of as a natural generalization of the spectral form factor to modular flow.  Such quantities are similarly expected to decay exponentially with $\Delta^2$.}  In particular, semiclassical states separated by  $\Delta\gg O(\sqrt{G})$ 
will be approximately orthogonal.  We can interpret this result as saying that (to good approximation) $s$ is a  semiclassical observable in our context, and that it has corresponding approximate eigenstates $|\psi(s)\rangle$ which are nearly orthogonal.    For the interested reader, Appendix~\ref{sec:JT} provides an explicit computation illustrating the analogue of this result in pure JT gravity, and which shows that the exponential decay is associated with complex saddle points. See especially the discussion around \Eqref{eq:integral}.

\begin{figure}
    \centering
    \includegraphics[scale=0.5]{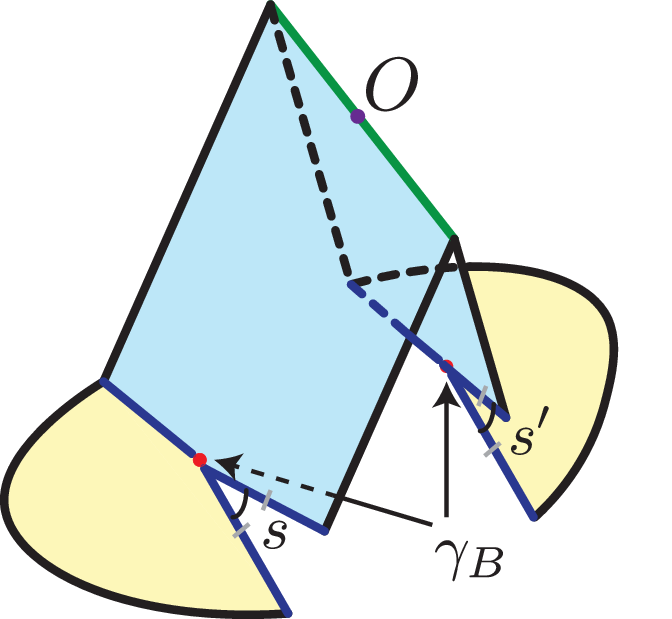}
    \caption{A general term in \Eqref{eq:cor} is computed using the Schwinger-Keldysh geometry shown. For $\Delta = (s-s')/2 \neq 0$, the two Lorentzian conical singularities have different strengths and the geometry is generally complex.  }
    \label{fig:off}
\end{figure}

Since $\tilde f(\bar s - \Delta), \tilde f^*(\bar s + \Delta)$ vary slowly with $\Delta$, the integral over $\Delta$ will be dominated by a saddle very close to $\Delta =0$.  Setting $\Delta =0$ in \Eqref{eq:cor} defines a diagonal approximation that yields a simple picture for the bulk state as a statistical mixture of the states $\ket{\psi(s)}$. Note that if $\tilde{f}(s)$ was sharply peaked at $s=s_0$, then \Eqref{eq:cor} would instead be well approximated by a single spacetime as expected for a semiclassical squeezed state.

We can now consider the reduced density matrix on subregion $A$ in the state $\ket{\psi_{\cA_0}}$.
It is given by
\begin{equation}\label{eq:rhoA}
    \rho_{A,\cA_0} = |N'|^2 \int ds\, ds' \tilde{f}(s) \tilde{f}^*(s')\tr_{\bar{A}}\(\ket{\psi(s)}\bra{\psi(s')}\).
\end{equation}
In terms of $\{\bar{s},\Delta\}$, the kernel $\tilde{f}(s)\tilde{f}^*(s')=\exp\[-\sigma^2 \((\bar{s}-s_0)^2+\Delta^2\) +2i x_0 \Delta\]$, and its magnitude decays slowly in both the $\bar s$ and $\Delta$ directions.

We can now take this density matrix and try to compute its $n$th moments for integer $n$ using the gravitational path integral.
For $n=1$, we have
\begin{equation}\label{eq:n1}
    \tr\(\rho_{A,\cA_0}\) = |N'|^2\int 2 d\bar{s}\,d\Delta\,\tilde{f}(\bar{s}+\Delta) \tilde{f}^*(\bar{s}-\Delta)\bra{\psi(\bar{s}-\Delta)}\ket{\psi(\bar{s}+\Delta)}.
\end{equation}
As discussed above, the magnitude $|\bra{\psi(\bar{s}-\Delta)}\ket{\psi(\bar{s}+\Delta)}|$ of the overlap $\bra{\psi(\bar{s}-\Delta)}\ket{\psi(\bar{s}+\Delta)}$ is strongly localized at $\Delta=0$. This suggests that the integral over $\Delta$ can be evaluated in the saddle point approximation, though we must also take into account that there will be a rapidly-varying phase set by the most-likely value $x_\psi$ of $x_B = \frac{\cA(\gamma_B)}{4G}$ in the original state $|\psi\rangle$.   The factor $\tilde{f}(\bar{s}+\Delta) \tilde{f}^*(\bar{s}-\Delta)$ also contributes a rapidly-varying phase that is of the opposite sign, and which is set by the parameter $x_0$ in \Eqref{eq:pseudoproj}.   For the moment, we will simply choose $x_0$ so that these phases cancel.  This is equivalent to taking the window function $f(\cA(\gamma_B))$ to be peaked at the area of the $\gamma_B$ in the classical geometry $g_\psi$ associated with $|\psi\rangle$.

On the other hand, the $\bar{s}$ direction is a soft mode over which we must still integrate.  
For integer $n>1$, to the extent that we can ignore any possible breaking of replica symmetry, we can use the Lewkowycz-Maldacena method \cite{Lewkowycz:2013nqa} for every choice of $\{\bar{s},\Delta\}$ to reduce the problem to the computation of the action of a solution with a cosmic brane of tension $T_n=\frac{n-1}{4nG}$ in the gravitational path integral satisfying the same boundary conditions as the calculation for $n=1$.
This yields\footnote{One might expect \Eqref{eq:n2} to integrate over $n$ copies of $\bar s, \Delta$ but in the saddle point approximation it is sufficient to integrate over one copy as in \Eqref{eq:n2} at the cost of neglecting subleading $O\(\log G\)$ contributions to the R\'enyi entropy.\label{footnote:saddle}}
\begin{equation}\label{eq:n2}
    \tr\(\rho_{A,\cA_0}^n\) \approx |N'|^{2n}\int 2 d\bar{s}\,d\Delta\,\tilde{f}(\bar{s}+\Delta)^n \tilde{f}^*(\bar{s}-\Delta)^n \bra{\psi(\bar{s}-\Delta)}O_n\ket{\psi(\bar{s}+\Delta)}^n,
\end{equation}
where $O_n$ represents the operator $e^{-\frac{n-1}{n} \frac{\cA(\gamma_A)}{4G}}$ inserting the cosmic brane of tension $T_n$.
We will in particular consider the case $n\approx1$ where we can neglect the backreaction caused by the cosmic brane.
We will now use the semiclassical approximation to  rewrite the matrix element $\bra{\psi(\bar{s}-\Delta)}O_n\ket{\psi(\bar{s}+\Delta)}$ by evaluating $O_n$ on the (for $\Delta \neq 0$, complex-valued) saddle-point geometry for the gravitational path integral that computes the overlap $\bra{\psi(\bar{s}-\Delta)}\ket{\psi(\bar{s}+\Delta)}$.  
This gives
\begin{equation}
\label{eq:mes}
\bra{\psi(\bar{s}-\Delta)}O_n\ket{\psi(\bar{s}+\Delta)} \approx
e^{-\frac{n-1}{n}\frac{\cA(\gamma_A(\bar{s},\Delta))}{4G}} \bra{\psi(\bar{s}-\Delta)}\ket{\psi(\bar{s}+\Delta)}, 
\end{equation}
where $\gamma_A(\bar{s},\Delta)$ is the area of the HRT surface for $A$ in the saddle corresponding to fixed $\bar{s},\Delta$ (which, as discussed above, will be complex for $\Delta \neq 0$).  

Inserting \eqref{eq:mes} into \eqref{eq:n2} gives
\begin{equation}\label{eq:n3}
    \tr\(\rho_{A,\cA_0}^n\) \approx |N'|^{2n}\int 2 d\bar{s}\,d\Delta\,\tilde{f}(\bar{s}+\Delta)^n \tilde{f}^*(\bar{s}-\Delta)^n e^{-(n-1)\frac{\cA(\gamma_A(\bar{s},\Delta))}{4G}} \bra{\psi(\bar{s}-\Delta)}\ket{\psi(\bar{s}+\Delta)}^n.
\end{equation}
There will again be a sharp decay in the $\Delta$ direction due to the overlap $\bra{\psi(\bar{s}-\Delta)}\ket{\psi(\bar{s}+\Delta)}^n$ while the factor $\tilde{f}(\bar{s}+\Delta)^n \tilde{f}^*(\bar{s}-\Delta)^n$ has a slowly-varying magnitude.  Thus we may again use saddle point methods to set $\Delta =0$, leaving us only with the integral over $\bar s$.  We will write $\gamma_A(\bar{s},\Delta=0)$ simply as $\gamma_A(\bar{s})$ below.

When $n=1+\epsilon$ with $\epsilon=O(1)$, we can use a further saddle point approximation to evaluate \Eqref{eq:n3} since the soft mode is then sharply localized by the contribution $\cA(\gamma_A(\bar s))$ from the cosmic brane action.  In particular, since $1/\sigma$ is large we will assume that  
$\cA(\gamma_A(\bar s))$ is minimized at some $\bar{s}_{\text{min}}$ such that $|\bar{s}_{\text{min}} - s_0| \ll  1/\sigma$ .    We may thus write
\begin{equation}\label{eq:n4}
    \tr\(\rho_{A,\cA_0}^n\) \approx |N'|^{2n} \tilde{f}(\bar{s}_{\text{min}})^n \tilde{f}^*(\bar{s}_{\text{min}})^n e^{-(n-1)\frac{\cA(\gamma_A(\bar{s}_{\text{min}}))}{4G}}.
\end{equation}
Note that this is a general result relating the R\'enyi entropy of $\rho_{A,\cA_0}$ to the area of $\gamma_A(\bar{s}_{\text{min}})$.

We now connect the above result to the constrained HRT surface $\gamma_{A:B}$. To do so, we specialize to the case  $\gamma_{A:B}= \check \gamma_{A:B}$ where $\gamma^B_{A:B}$ intersects $\gamma^{\bar B}_{A:B}$ at a constant boost angle $\check s$.
We then claim that $\bar s_{\text{min}} = -\check s$, and that the area of $\gamma_{A}(\bar s_{\text{min}})$ is precisely that of $\check \gamma_{A:B}$. Furthermore, in a given classical geometry the rapidity $\check s$ will be independent of $G$.  As a result, since $1/\sigma$ is parametrically large, we will indeed find $|\bar s_{\text{min}} -s_0|\ll 1/\sigma$.

To show this, we first prove that $\cA(\check \gamma_{A:B})$ gives the area of the HRT surface $\gamma_A(-\check s)$ in the geometry $g_{\psi(-\check s)}$.  This $g_{\psi(-\check s)}$ is by definition the result of applying a BCP-kink transformation with boost parameter $-\check s$ to $g_\psi$.  Since this does not change either the entanglement wedge of $B$ or that of $\bar B$, the spacetime $g_{\psi(-\check s)}$ contains extremal surfaces corresponding to both $\gamma^B_{A:B}$ and $\gamma^{\bar B}_{A:B}$ that are anchored both to the boundary and to a common locus on $\gamma_B$.  Furthermore, due to the relative boost by $-\check s$, their tangents are now continuous across $\gamma_B$.  And since extremal surfaces satisfy a second order partial differential equation, it follows that they in fact join smoothly across $\gamma_B$ to form a surface that is everywhere extremal.  This surface also satisfies the same homology constraint as $\gamma_A(-\check s)$ and, indeed, it must be the minimal such surface on some Cauchy slice $\Sigma$ that contains $\gamma_B$ (since our maximin construction ensured that this was the case for the original $\gamma_{A:B}$).  As a result, the representative argument of Sec.~3.1 of Ref.~\cite{Wall:2012uf} again shows that, in a spacetime satisfying the null energy condition, there can be no smaller  extremal surface that is both anchored and homologous to $A$. 

In particular, when the maximin surface is unique we see that the original constrained HRT surface $\gamma_{A:B}$ is just the BCP-kink transform of the HRT surface $\gamma_A(-\check{s})$ in the spacetime $g_{\psi(-\check s)}$.
 It now follows immediately from \Eqref{eq:cHRTlb} that the choice $\bar s = -\check s$ minimizes the area of the surfaces $\gamma_A(\bar s)$ as claimed, and thus that
\begin{equation}
\label{eq:final}
	S_{1+\epsilon}(\rho_{A,\cA_0})\approx \frac{\cA(\check \gamma_{A:B})}{4G},
\end{equation}
where we have ignored logarithmic corrections arising from the prefactors in \Eqref{eq:n4}.

Thus far we have considered only the case where the window function $f(\cA(\gamma_B))$ is peaked at the area of the surface $\gamma_B$ in the classical geometry $g_\psi$ associated with the state $|\psi\rangle$.  However, the case of more general window functions is now straightforward.  We may simply choose the operator $\Pi_{\cA_0}$ to be a product of two separate (commuting) window operators.  We take the first to be of the form \eqref{eq:pseudoproj} with $s_0=0$, some choice of $x_0$, and a width $\sigma_1$ of order $G^{-\alpha_1}$ for $1/2>\alpha_1 >0$.  Acting with the associated $\Pi_{\cA_0,1}$ thus produces what Ref.~\cite{Dong:2022ilf} called a squeezed fixed-area-state $|\psi_1\rangle := \Pi_{\cA_0,1}|\psi\rangle$ with area $\cA_0$. We then take the second window operator $\Pi_{\cA_0,2}$ to again be of the form \eqref{eq:pseudoproj}, with the same $x_0$, but now with arbitrary $s_0$ (that is independent of $G$) and with a width $\sigma_2$ of order $G^{\alpha_2}$  for some $\alpha_2 > 0$.  Note that the product 
$\Pi_{\cA_0}= \Pi_{\cA_0,2}\Pi_{\cA_0,1}$ is again of the form \eqref{eq:pseudoproj} with the above values of $x$, $s_0$ and with a width slightly smaller than $\sigma_2.$  Thus this final operator $\Pi_{\cA_0}$ prepares the desired pseudo-eigenstate.

The point of this two-step construction is that the squeezed state $|\psi_1\rangle$ is still semiclassical and, in particular, it is still well-approximated by a single classical geometry of the form described in Ref.~\cite{Dong:2022ilf}.  We emphasize that this geometry solves the usual classical equations of motion without additional sources.  In particular, while the Euclidean (or complex) saddle associated with a path integral definition of this state will have a conical singularity at $\gamma_B$, there is no conical singularity in the Lorentzian solution.  While the Lorentzian solution will generally contain power law divergences near the null congruences orthogonal to $\gamma_B$, these satisfy the equations of motion; see Ref.~\cite{Dong:2022ilf} for details.

Since $|\psi_1\rangle$ is semiclassical, and since it is now peaked at a value of $\cA(\gamma_B)$ that is very close to $4G x_0$, we can analyze the pseudo-eigenstate $\Pi_{\cA_0} |\psi\rangle= \Pi_{\cA_0,2}|\psi_1\rangle$ by repeating the argument given above involving Eqs.~\eqref{eq:cor}--\eqref{eq:final} replacing the state $|\psi\rangle$ by  $|\psi_1\rangle$ and $\Pi_{\cA_0}$ by $\Pi_{\cA_0,2}$.  The addition of power law divergences does not significantly affect the discussion.  In particular, since we are focused here on Einstein-Hilbert gravity, the areas of codimension-2 surfaces remain finite and well-defined (see Ref.~\cite{Dong:2022ilf}). 
And while the most likely value $x_1$ of $\hat x_B$ in $|\psi_1\rangle$ will differ slightly from $x_0$, one may check that the difference is sufficiently small that setting $\Delta=0$ in \Eqref{eq:n3} remains a good approximation to the saddle-point result.\footnote{Or, alternatively, one can adjust the peak of the window function for $\Pi_{\cA_0,1}$ so as to place the peak of $|\psi_1\rangle$ at the desired value $x_0$.}  We thus again arrive at the desired result \eqref{eq:final}.

\subsection{Density matrix interpretation}

We will now interpret the above results in terms of the structure of the reduced density matrix \eqref{eq:rhoA} on subregion $A$.  In doing so, we will see that much of the above story is reproduced by applying standard bounds on R\'enyi and von Neumann entropies.  

We will take as input two lessons from the discussion above.  
The first lesson is that the density matrix \eqref{eq:rhoA} is nearly diagonal in $s$.  This fact is in strong accord with the expectation that in the bulk semiclassical limit, $s$ describes a position-independent boost angle at $\gamma_B$. As a result, in this limit it can be measured locally in both the entanglement wedge of $A$ and in that of $\bar A$.  The second lesson, however, is that this diagonality is only approximate.  Indeed, it arises {\it only} in the semiclassical limit. As a result, the quantum state $\ket{\psi}$ has an $O(\sqrt{G})$ width in the boost angle $s_B$. This width is then shared by each $|\psi(s)\rangle$, so that the $s,s'$ terms in the density matrix \eqref{eq:rhoA} for subregion $A$ superpose coherently over any $s$-interval with width smaller than $O(\sqrt{G})$.  We therefore expect that \Eqref{eq:rhoA} can be approximated by choosing a discrete set ${\cal S}$ of $\bar s$-values and considering the state $\tilde \rho_{A,\cA_0}$ below:
\begin{equation}\label{eq:block}
\rho_{A,\cA_0} \approx \tilde \rho_{A,\cA_0} := \sum_{\bar{s} \in {\cal S}} c(\bar s) |\tilde{f}(\bar{s})|^2 \tilde \rho_{A,\bar{s}},
\end{equation}
where $\tilde \rho_{A,\bar{s}}$ is a reduced density matrix on subregion $A$ obtained  from \Eqref{eq:rhoA} by inserting an additional Gaussian window function $e^{-\left[(s - \bar s)^2 + (s'-\bar s)^2\right]/2\alpha^2}$ of width $\alpha \sim \sqrt{G}$ and normalizing the result.\footnote{In \Eqref{eq:block} we did not include contributions from slightly off-diagonal terms (i.e., from $s$, $s'$ that are within a few standard deviations of each other). Including these would contribute a subleading $O\(\log G\)$ term to the R\'enyi entropy, similar to what was explained in footnote \ref{footnote:saddle}.}  The factor $c(\bar s)$ is a slowly-varying positive function defined by a combination of the factor involved in the above renormalization of $ \tilde \rho_{A,\bar{s}}$ and the local density of $\bar s$ values in ${\cal S}$.  What is important for us is that $c(\bar s)$ will again be polynomial in $G$. The coefficients $c(\bar{s}) |\tilde{f}(\bar{s})|^2$ play the role of a probability distribution over the boost parameter $\bar{s}$.

The ansatz \eqref{eq:block} describes the fixed-area reduced density matrix as a mixture of reduced density matrices on subregion $A$ associated with different values of the flow parameter $\bar s$.
Note that we have not required the density matrices $\tilde \rho_{A,\bar{s}}$ to be orthogonal to each other in any way.  Non-zero overlaps are allowed, and we will account for such overlaps in our analysis below.  We expect the approximation \eqref{eq:block} to be accurate in the limit where the set ${\cal S}$ becomes dense, though for later use we will assume that the number of points in ${\cal S}$ is taken to be at most polynomially large in $1/G$.

The width $\alpha \sim O(\sqrt{G})$ is optimal since making it much sharper or wider would typically reduce the width of either $\cA_B$ or $s_B$.  Since neither of these quantities commutes with $\cA(\gamma_A)$, either case would make it difficult to have a semiclassical formula for entropies of $A$.

We can now use \Eqref{eq:block} to compute R\'enyi entropies of $A$.
In particular, for a mixture of density matrices $\rho=\sum_i p_i \rho_i$, we have the inequality
\begin{equation}\label{eq:ineq1}
    \frac{1}{1-n}\log(\sum_{i} p_i e^{-(n-1)S_n(\rho_i)}) \leq S_n(\rho) \leq \frac{1}{1-n} \log(\sum_{i} p_i^n e^{-(n-1)S_n(\rho_i)}),
\end{equation}
which follows from Schur concavity of the R\'enyi entropy (see Eq.(6.51) of Ref.~\cite{nielsen}).
Applying the above inequality to \Eqref{eq:block}, we have
\begin{equation}\label{eq:ineq1}
    \frac{1}{1-n}\log(\sum_{\bar{s}} c(\bar s)|\tilde f({\bar{s}})|^2 e^{-(n-1)S_n(\tilde\rho_{A,{\bar{s}}})}) \leq S_n(\tilde \rho_{A,\cA_0}) \leq \frac{1}{1-n} \log(\sum_{\bar{s}} c(\bar s)^n |\tilde f({\bar{s}})|^{2n} e^{-(n-1)S_n(\tilde\rho_{A,\bar{s}})}),
\end{equation}
We emphasize that the bounds hold certainly for the density matrix $\tilde \rho_{A,\cA_0}$, but that additional so-far-uncontrolled errors can enter if we wish to interpret \Eqref{eq:ineq1} as placing bounds on the original density matrix $\rho_{A,\cA_0}$.  Thus the current discussion represents an interpretation of our earlier results, in which the approximation of $\rho_{A,\cA_0}$ by \Eqref{eq:block} will be justified a posteriori by using \Eqref{eq:ineq1} to reproduce \Eqref{eq:final}. 

Since each individual density matrix $\tilde\rho_{A,\bar{s}}$ has $O(\sqrt{G})$ fluctuations, it is natural to expect that the associated von Neumann entropy can be analyzed a la Lewkowycz-Maldacena \cite{Lewkowycz:2013nqa}, and thus that for $n=1+\epsilon$, the R\'enyi entropies are computed by a cosmic brane prescription and that  backreaction from the brane gives only an $O(\frac{\epsilon}{G})$ correction.  Thus we write 
\begin{equation}
    S_n(\tilde\rho_{A,\bar{s}}) = \frac{\cA(\gamma_{A,\bar{s}})}{4G} + O\(\frac{\epsilon}{G}\),
\end{equation}
where $\gamma_{A,\bar{s}}$ is the HRT surface in the classical geometry $g_{\psi(\bar s)}$ corresponding to $\ket{\psi(\bar{s})}$.
The sums in \Eqref{eq:ineq1} involve a slowly varying factor (from $c(\bar s)|\tilde f(\bar s)|^2$) multiplied by a factor $\exp\[-\frac{\epsilon \gamma_{A,\bar{s}}}{4G}\]$  whose variation is parametrically more rapid.
Thus, the sum is dominated by the term with the smallest area which, as discussed before, is given by $\cA(\check \gamma_{A:B})$ in the case where $\gamma^B_{A:B}$, $\gamma^{\bar B}_{A:B}$ meet at a constant boost angle (which we now assume).

Approximating the sums  in \Eqref{eq:ineq1} by integrals and using the fact that $\tilde f(\bar s)$ is Gaussian, we find that the LHS and RHS differ by two kinds of terms: an $O\(\frac{\sigma^2}{\epsilon}\)$ difference arising from the shift in the peak of the Gaussian and $O(\log G)$ corrections from the prefactors.
Thus, putting everything together we have
\begin{align}\label{eq:cross}
	S_{1+\epsilon}(\rho_{A,\cA_0})&= \frac{\cA(\check \gamma_{A:B})}{4G} +O\(\frac{\epsilon}{G}\)+O\(\frac{\sigma^2}{\epsilon}\)+O\(\log G\).
\end{align}
In particular, we see that while the errors are small for $\epsilon\gg\sigma^2$, the errors become large in the strict $\epsilon\to0$ limit required to compute the von Neumann entropy.   This 
non-commutativity of  the $n\to 1$ and $G\to 0$ limits is analogous to that described in 
\cite{Akers:2020pmf} associated with computations of entropy near an HRT-phase transition.

To compute the von Neumann entropy we can use the analogous inequality \cite{nielsen2001quantum}
\begin{equation}
    \sum_{\bar{s}} c(\bar s) |\tilde{f}(\bar{s})|^2 S(\tilde\rho_{A,\bar{s}}) \leq S(\tilde\rho_{A,\cA_0}) \leq \sum_{\bar{s}} c(\bar s) |\tilde{f}(\bar{s})|^2 S(\tilde\rho_{A,\bar{s}}) - \sum_{\bar{s}} c(\bar s)|\tilde{f}(\bar{s})|^2 \log \[c(\bar s) |\tilde{f}(\bar{s})|^2\].
\end{equation}
The LHS and RHS  differ by a Shannon term which is bounded above by $\log k$ where $k$ is the number of relevant terms in the sum.
For our case, we can truncate the sum to the region $|\bar{s}-s_0|\leq \frac{1}{\sigma}$.
Since the size of the set ${\cal S}$ of $\bar s$ values used in the sum in \Eqref{eq:block} was taken to be at most polynomial in $1/G$, this Shannon term is at most $O(\log G)$.  Note that, since $\sigma$ is at most polynomial in $1/G$, our restriction on the size of ${\cal S}$ still allows an arbitrary polynomially-large density of $\bar s$ values.
Thus, we have
\begin{equation}
\label{eq:vNfrombounds}
    S(\rho_{A,\cA_0}) = \sum_{\bar{s}} c(\bar s) |\tilde{f}(\bar{s})|^2 S(\tilde\rho_{A,\bar{s}}) + O(\log G).
\end{equation}
Since we expect $S(\tilde\rho_{A,\bar{s}})$ to be given by an HRT-area as described above, the result \eqref{eq:vNfrombounds} can be interpreted as the expectation value of this area (up to $O(\log G)$ corrections).
This can be thought of as the application of the HRT formula to a superposition of spacetimes with a widely spread probability distribution \cite{Almheiri:2016blp,Harlow:2016vwg,Akers:2018fow}.

To summarize, the key point in the above calculation is that the state \eqref{eq:projstate} has a soft mode associated with the boost angle across the HRT surface.
When evaluating the $n\approx 1$ R\'enyi entropy, the soft mode is stabilized by the dependence of $\cA(\gamma_A (\bar s))/4G$ on this boost angle. As a result, the integral can be evaluated using a saddle point approximation at the smallest value of this area.
However, for the entanglement entropy, we need to perform the full integral over the soft mode. The final answer is thus {\it not} given by the area of a single surface in a single geometry.

A final important point is the dependence of the above calculation on the area width $\sigma$ and the parameter $s_0$.
In calculating the $n\approx 1$ R\'enyi entropy we have assumed that $\bar{s}=-\check s$ is within a few standard deviations from the maximum $s_0$ in the probability distribution for $s$.  This is a natural assumption in the limit where the distribution in $s$ becomes broad.
However, if we choose the original width $\sigma$ of $\cA_B/4G$ to be larger than $O(1)$, then the width in $s$ is small and, in particular, is likely to be small compared with the separation between $s_0$ and $-\check s$. In that case we are considering squeezed states for which case the Gaussian distribution over $\bar{s}$ remains sharp in the semiclassical limit and we instead obtain the usual HRT formula for the spacetime associated with the boost $\bar s = s_0$.
On the other hand, if we choose the width $\sigma$ to be {\it non-perturbatively} small in $G$ then, even though the distribution in $s$ will in some sense be broad enough to access the regime near $-\check s$, the fact that we may then be sensitive to the discreteness of area-eigenvalues implies that there can be no good notion of a conjugate variable $s$.  In addition, the $O(\log G)$ error terms in \Eqref{eq:vNfrombounds}  become very large and it is difficult to claim any precise result for the entropies.  (Note that a similar comment applies to the logarithmic corrections to \Eqref{eq:final} from the neglected prefactors in \Eqref{eq:n4}.)
This is natural since  such a regime is far from being semiclassical.
What is interesting, then, is that there is an intermediate regime where $\sigma$ is polynomially small in $G$ that yields a broad superposition over spacetimes with different $\bar s$-values for which the $n\approx 1$ R\'enyi entropies can still be obtained from a semiclassical computation in a single spacetime.
This is in agreement with the expectation of Ref.~\cite{Almheiri:2016blp} that the area operator behaves linearly for a small superposition over spacetimes.

\section{Discussion}
\label{sec:disc}

The above work considered the constrained HRT surfaces defined in Ref.~\cite{Chen:2018rgz} and found an entropic interpretation for their areas in situations where their failure to be smooth is defined by a single constant boost angle.
This is closely related to the results of Ref.~\cite{Chen:2018rgz} which interpreted modular flow as a disentangler.
We made use of the fact that a fixed-area state defined by a boundary subregion $B$ automatically explores a wide range of BCP kink transformations.
The action of modular flow on a semiclassical geometry is well approximated by the BCP kink transformation \cite{jlms}.
Thus, a fixed-area state effectively explores a wide range of modular flows applied to the original semiclassical state from which the fixed-area state was obtained.
Consequently, computations of the $n\approx 1$ R\'enyi entropy for another boundary subregion $A$ are dominated by values of the flow parameter that minimize the area $\cA (\gamma_A)$ of the HRT surface $\gamma_A$. Interestingly, this differs from the behavior of the von Neumann entropy $S(A)$, which is one of the qualitative differences between our entropic interpretation and that of Ref.~\cite{Chen:2018rgz}.

As a small technical comment, our introduction of a possible non-trivial oscillatory phase $\exp{-is_0\cA_B/4G}$ in the window operator $\Pi_{\cA_0}$ used to construct the fixed-area state \eqref{eq:projstate} generalizes the class of fixed-area states that have been explicitly described in the past.  In particular, in the analysis of Lorentzian geometries for squeezed states (so that $\sigma$ is no smaller than $O(1)$) performed in Ref.~\cite{Dong:2022ilf}, our $s_0$ was implicitly set to zero.  However, including the effect of $s_0$ was not difficult, especially since it was noted in \secref{sec:entropic} that this phase constitutes a source for a Lorentz-signature conical defect as described in \figref{fig:lor}.

We will now comment on various possible corrections to \Eqref{eq:final}. In the case where the classical geometry $g_\psi$ of $|\psi\rangle$ is smooth and we apply a window operator $\Pi_{\cA_0}$ with $\cA_0$ equal to the area $\cA(\gamma_B)$ in $g_\psi$, 
higher derivative corrections to the bulk Einstein-Hilbert action will be straightforward to include if the corrected geometric entropies continue to generate BCP kink transformations.  As will be discussed in Ref.~\cite{geom}, this is the case for some class of higher derivative theories.  Further issues appear at general values of $\cA_0$, as the associated power law divergences become more singular 
in the presence of higher-derivative corrections.  It is not clear how to accommodate this in our calculation, though it is plausible that an analogous result would follow from a regulated calculation in which one smears out the fixed-area surface over a codimension-zero region.  In that context, there should be no divergences.

In attempting to include quantum corrections, a natural extension of our proposal would be to replace fixed-area states by states where we restrict the boundary modular Hamiltonian to a small window of eigenvalues.  Such a restricted state can then be expressed as a superposition over modular flowed states.
However, the precise bulk dual of such states is not well understood since restricting the bulk modular Hamiltonian would require non-local sources acting on the entire entanglement wedge, unlike the localized conical defect source for fixed-area states.
It would be interesting to understand this better in the future as well.

We would also like to discuss the limitations of our analysis.
We were able to provide an entropic interpretation for $\gamma_{A:B}$ only when it intersects $\gamma_B$ at a constant boost angle.
As pointed out in Ref.~\cite{Chen:2018rgz}, this case is not generic.  
Even in cases with two boundary dimensions, it can fail when the region $A$ contains multiple disconnected intervals.

It would thus be interesting to find a generalization of fixed-area states that would allow one to explore position-dependent boosts, thus opening up the possibility of a more general entropic interpretation for constrained HRT surfaces. Since the HRT area $\frac{A_\gamma}{4G} = \frac{1}{4G}\int_\gamma \sqrt{h}$ (with $h$ the determinant of the induced metric on $\gamma$) generates a constant boost around $\gamma$, it may be tempting to suppose that operators of the form ${\Phi}_\gamma = \frac{1}{4G}\int_\gamma \sqrt{h} \phi$ generate position-dependent boosts of local rapidity proportional to the scalar field $\phi$, and thus such operators would thus lead directly to the generalization desired.  Indeed, had we considered a context in which $\gamma$ was a spacetime boundary (at which the metric was determined by boundary conditions), then it seems likely that this would be the case.  However, in the case where $\gamma$ is determined dynamically, a critical step in the analysis of \cite{Kaplan:2022orm} was to observe that (by definition) the functional $\frac{A_\gamma}{4G}$ is stationary with respect to variations of the HRT surface $\gamma$.  This is also closely related to the observation of \cite{Bousso:2020yxi} that a BCP kink transformation applied around a non-extremal surface $\gamma$ generally leads to violations of the constraints.  In the same way, we should expect ${\Phi}_\gamma = \frac{1}{4G}\int_\gamma \sqrt{h} \phi$ to generate position-dependent boosts around $\gamma$ only when $\gamma$ extremizes the functional ${\Phi}_\gamma.$  It is thus far from clear how such operators can be useful in general.

Our results also have implications for attempts to understand the random tensor networks of Ref.~\cite{Hayden:2016cfa} as models of general holographic states.  Such networks have a flat R\'enyi spectrum for each boundary subregion.  For any fixed subregion $B$, it is thus natural to associate such networks with bulk states that have a fixed area for  $\gamma_B$ \cite{Akers:2018fow,Dong:2018seb}.  In gravity, due to a lack of commutativity of $\cA (\gamma_B)$ with $\cA (\gamma_A)$ for other boundary subregions $A$, it is difficult to fix many such areas simultaneously \cite{Bao:2018pvs}.  One might therefore try to proceed by studying alternative operators designed to better commute with $\cA (\gamma_B)$; see also related discussion in Ref.~\cite{examples}.  But we see here that doing so with $\cA(\gamma_{A:B})$ can lead to new issues, such as a failure for the $n\rightarrow 1$ and $G\rightarrow 0$ limits to commute, with the result that
the new operator $\cA(\gamma_{A:B})$ fails to represent a von Neumann entropy after fixing the area of $\gamma_B$.

Finally, we note that the methods used in this paper can also be applied to geodesics in arbitrary dimensions that compute correlation functions in fixed-area states.
For instance, consider the microcanonical thermofield double state \cite{Marolf:2018ldl}, where the energy has been fixed to some polynomially small width in $G$.
The left-right two-point correlation function of heavy operators in this state would be similarly computed by a constrained geodesic intersecting the bifurcate horizon by arguments similar to those made here.
This comment is closely related to the ideas of Ref.~\cite{Faulkner:2018faa}.

\acknowledgments

We would like to thank Chris Akers, Luca Iliesiu, Geoff Penington and Arvin Shahbazi-Moghaddam for useful discussions. PR is supported in part by a grant from the Simons Foundation, and by funds from UCSB. This material is based upon work supported by the Air Force Office of Scientific Research under award number FA9550-19-1-0360.
This work was supported in part by the Berkeley Center for Theoretical Physics; by the Department of Energy, Office of Science, Office of High Energy Physics under QuantISED Award DE-SC0019380 and under contract DE-AC02-05CH11231. PR was supported by the National Science Foundation under Award Number 2112880.

\appendix

\section{JT model}
\label{sec:JT}

In this appendix, we consider pure JT gravity.  This theory was canonically quantized in Ref.~\cite{Harlow:2018tqv}.
Equivalently, it can be described by the gravitational path integral restricted to the disk topology \cite{Yang:2018gdb}.
The phase space variables can be chosen to be $\phi_h$, the value of the dilaton at the horizon, and the conjugate variable $s$ which corresponds to the boost across the horizon.
Another useful phase space variable is the renormalized geodesic length $L$ between the two boundaries.

We will treat $\phi_h$ below analogously to the HRT area of $B$ discussed in the main text.  We will also take $L$ to be an analogue of the HRT area of $A$.
Of course, $L$ has no entropic interpretation in 2D gravity but it may still serve as a useful model for higher dimensional contexts  where codimension-2 surfaces with an entropic interpretation can reach from boundary to boundary.

First we can define a fixed-area state using a Gaussian window function as
\begin{equation}
    \ket{\phi_0} = \int d\phi_h \,e^{-\frac{(\phi_h-\phi_0)^2}{2\sigma^2}}\ket{\phi_h}.
\end{equation}
The analogue of our main-text computation of R\'enyi entropies for subregion $A$  is then to consider a correlation function of the form
\begin{align}
    S_n(A) &= \bra{\phi_0}e^{-(n-1)\frac{L}{4\beta}}\ket{\phi_0}\\
    &= \int d\phi_h \,d\phi_h'\,dL\,e^{-(n-1)\frac{L}{4\beta}}e^{-\frac{(\phi_h-\phi_0)^2}{2\sigma^2}}e^{-\frac{(\phi_h'-\phi_0)^2}{2\sigma^2}}\psi_{\phi_h}(L)\,\psi^*_{\phi_h'}(L),
\end{align}
where we have used the fact that $\beta$ plays the role of $G$ in JT gravity and 
where $\psi_{\phi_h}(L) = \bra{L}\ket{\phi_h}$ is the wavefunction of a dilaton eigenstate in the $L$ basis.  We have also set $\phi_b=1$.

The wavefunctions were written down explicitly in Ref.~\cite{Harlow:2018tqv} and are given by Bessel functions.
However, for our purposes, it will be sufficient to mention only a few properties of $\psi_{\phi_h}(L)$.
The first is that $\psi_{\phi_h}(L)$ decays rapidly below the classical turning point $L_c$ which corresponds to the length of the geodesic on the time symmetric surface in the associated classical geometry. Thus, it can be taken to vanish for $L<L_c$. The second is that for $L>L_c$ this $\psi_{\phi_h}(L)$ is well-approximated by $e^{-2i\phi_h L}+ Re^{2i\phi_h L}$, where $R$ is a pure phase. 
This simple form is associated with the fact that
the conjugate variable to $\phi_h$ can be described by a relative boost $s$ between the left and right wedges, and also with the fact that (unless $s$ is very small), the length $L$ increases linearly with $|s|$.  In particular, the fact that there are two branches stems from the fact that relative boosts of $\pm s$ lead to the same value of $L$.

We will first perform the integrals over $\phi_h$ and $\phi_h'$.
Using the saddle point approximation, one finds that we are left with
\begin{equation}
    \bra{\phi_0}e^{-(n-1)L/4\beta}\ket{\phi_0} \approx\int_{L_c}^{\infty} dL \,|I(L)|^2 e^{-(n-1)L/4\beta},
\end{equation}
where $L_c=-2\log\(\phi_h\)$ is the smallest renormalized length within the classically allowed region and $I(L)$ is given by
\begin{equation}
    I(L)\approx e^{-2L^2 \sigma^2} \(e^{-2i L \phi_0}+R e^{2i L \phi_0}\).
\end{equation}
Thus, we see that we have a wide Gaussian that decays with width $\frac{1}{\sigma}$.
Finally, the term $e^{-(n-1)\frac{L}{4\beta}}$ for any $n-1=O(1)$ creates a pressure to pick up the smallest value of $L$ within the classically allowed region.
This is precisely the geodesic at the time symmetric slice which is the analogue of the constrained HRT surface.
On the other hand, we could also have evaluated the analog of the entanglement entropy $\bra{\phi_0}\frac{L}{4\beta}\ket{\phi_0}$ which would be given by the average value of $L$, instead of the length of a single geodesic.

In order to make a further connection with our calculation of interest, we can re-express the fixed-area state in terms of a superposition over BCP kink-transformed states.
The terms arising in the computation of the R\'enyi entropy then take the form $\bra{\psi(s)}e^{-\bar{\epsilon} L}\ket{\psi(s')}$ where $\bar{\epsilon}=\frac{n-1}{4\beta}$.
Using the thermofield double state for $\ket{\psi}$, one of the terms (obtained from incoming waves) takes the form
\begin{eqnarray}\label{eq:s1s2}
    \bra{\psi(s_1)}e^{-\bar{\epsilon} L}\ket{\psi(s_2)}&\supset& \int d\phi_1\,d\phi_2\, dL\, \exp\Bigl[ 2\pi \phi_1 -\frac{\beta \phi_1^2}{2}+2\pi \phi_2 -\frac{\beta \phi_2^2}{2}-2 i \phi_1 L + 2 i \phi_2 L  \cr &+& i \phi_1 s_1-i\phi_2 s_2 -\bar{\epsilon} L\Bigr],
\end{eqnarray}
where we have used the form of the Hartle-Hawking state.
Evaluating the integrals in the saddle point approximation, we have
\begin{eqnarray}\label{eq:integral}
    \bra{\psi(s_1)}e^{-\bar{\epsilon} L}\ket{\psi(s_2)} \supset \exp\[\frac{4\pi^2}{\beta}-\frac{1}{4}(s_1+s_2)\bar{\epsilon}+\frac{\beta \bar{\epsilon}^2}{16}-\frac{(s_1-s_2)^2}{4\beta}+\frac{2\pi i(s_1-s_2)}{\beta}\],
\end{eqnarray}
where we have used the complex saddle at $\phi_1=\frac{2 \pi}{\beta}+i\(\frac{s_1-s_2}{2\beta}+\frac{\bar{\epsilon}}{4}\)$, $\phi_2=\frac{2 \pi}{\beta}+i\(\frac{s_1-s_2}{2\beta}-\frac{\bar{\epsilon}}{4}\)$ and $L = \frac{s_1+s_2}{4}-\frac{\beta \bar{\epsilon}}{8}$.
It is easy to see that \Eqref{eq:integral} vanishes for $|s_1-s_2|\gg O(\sqrt{\beta})$ as claimed.
As discussed previously, $\beta$ plays the role of $G$ in this theory.
Moreover, the imaginary part in the exponent leads to wild oscillations when $\Delta := \frac{s_2-s_1}{2} \neq 0$.
Note that a similar term with the opposite phase arises in the computation of the R\'enyi entropy as well, and the overall integral is purely real.
We also see that there is a pressure to decrease the value of $\bar{s}=\frac{s_1+s_2}{2}$ until the classical turning point $L_c$.
The R\'enyi entropy is then obtained by the integral over $s_1,s_2$ with a wide Gaussian kernel.
In effect, we have shown that this calculation is dominated by the diagonal terms where $\Delta=0$ and at the value $\bar{s}=0$ corresponding to the constrained HRT surface.

In this approximation, one encounters another kind of ``interference" term where we use the part of the wavefunction containing the reflected wave for one of the terms.
We do not expect a saddle of this kind in general.
In the approximation we worked in, this term is given by
\begin{eqnarray}
\label{eq:intterm}
    \bra{\psi(s_1)}e^{-\bar{\epsilon} L}\ket{\psi(s_2)}&\supset& \int d\phi_1\,d\phi_2\, dL\, R \exp\Bigl[2\pi \phi_1 -\frac{\beta \phi_1^2}{2}+2\pi \phi_2 -\frac{\beta \phi_2^2}{2}
    \cr &+&2 i \phi_1 L + 2 i \phi_2 L +i \phi_1 s_1-i\phi_2 s_2 -\bar{\epsilon} L\Bigr].
\end{eqnarray}
One can check that the naive saddle for this integral is located far from the original contour, so that it is outside the regime of validity of our approximation.
Instead, a careful analysis involving the exact wavefunctions $\psi_{\phi_h}(L)$ would be required to determine if the original contour can be deformed so as to access this saddle in a useful way.  

However, we will follow a different route since we are only interested in working perturbatively around $\bar{\epsilon}=0$. 
At $\bar{\epsilon}=0$, one can do the exact calculation in \Eqref{eq:s1s2} without introducing a complete set of $L$ eigenstates.
One can then check that we obtain precisely the result of \Eqref{eq:integral} in the saddle point approximation at $\bar{\epsilon}=0$.
This shows that at $\bar{\epsilon}=0$, the interference term \eqref{eq:intterm} does not contribute.
Since we are working perturbatively in $\bar{\epsilon}$, this continues to be true at higher orders in $\bar{\epsilon}$.
In summary, we have explicitly demonstrated the diagonal approximation in JT gravity.

\addcontentsline{toc}{section}{References}
\bibliographystyle{JHEP}
\bibliography{references}

\end{document}